# A new design strategy based on a deterministic definition of the seismic input to overcome the limits of design procedures based on probabilistic approaches

Marco Fasan, Claudio Amadio, Salvatore Noè
*University of Trieste, Department of Engineering and Architecture - Via Alfonso Valerio 6/1, 34127 Trieste, Italy*

Giuliano Panza
*University of Trieste, Department of Mathematics and Geosciences - Via Weiss 4, 34127 Trieste, Italy*
*The Abdus Salam International Centre for Theoretical Physics ICTP, SAND Group – 34151 Strada Costiera 11, Trieste, Italy*
*China Earthquake Administration, Institute of Geophysics, Minzudaxue Nanlu 5, Haidian District, 100081 Beijing, P.R. China*
*International Seismic Safety Organization (ISSO) - www.issoquake.org*

Andrea Magrin, Fabio Romanelli, Franco Vaccari
*University of Trieste, Department of Mathematics and Geosciences - Via Weiss 4, 34127 Trieste, Italy*
*The Abdus Salam International Centre for Theoretical Physics ICTP, SAND Group – 34151 Strada Costiera 11, Trieste, Italy*

ABSTRACT
In this paper, a new seismic Performance Based Design (PBD) process based on a deterministic definition of the seismic input is presented. The proposed procedure aims to address the following considerations, arisen from the analysis of seismic phenomena, which cannot be taken in account using standard probabilistic seismic input (PSHA): a) any structure at a given location, regardless of its importance, is subject to the same shaking as a result of a given earthquake, b) it is impossible to determine when a future earthquake of a given intensity/magnitude will occur, c) insufficient data are available to develop reliable statistics with regards to earthquakes. On the basis of these considerations, the seismic input at a given site - determined on the basis of the seismic history, the seismogenic zones and the seismogenic nodes - is defined using the Neo Deterministic Seismic Hazard Assessment (NDSHA). Two different analysis are carried out at different levels of detail. The first one (RSA) provides the "Maximum Deterministic Seismic Input" as a response spectra at the bedrock ($MDSI_{BD}$), similarly to what is proposed by the codes. The second one (SSA) takes the site effects into account, providing a site specific seismic input ($MDSI_{SS}$). A SSA provides realistic site specific seismograms that could be used to run time history analysis even where no registrations are available. Reviewing the standard PBD procedure, $MDSI_{SS}$ is always associated with the worst structural performance acceptable for a building, called Target Performance Level (TPL). In this way, the importance of the structure (risk category) is taken into account by changing the structural performance level to check rather than to change the seismic input.

## 1 INTRODUCTION

Nowadays, almost all seismic building codes are based on principles of Performance Based Design (PBD). The main contribution to the development of this philosophy of design has been given by the Vision 2000 report (SEAOC 1995). This document, and the following papers, define a series of performances (in terms of acceptable damage) that a building should reach during earthquakes of different strength. These performance levels are usually defined as: Operational Limit (OL), Immediate Occupancy (IO), Life Safety (LS) and Collapse Prevention (CP), (ASCE 2013, CEN 2005, C.S.LL.PP. 2008). The earthquake input related to each of them (the earthquake strength) is chosen as a function of a fixed probability of exceeding it ($P_{EY}$) in a range of time $Y$ (reference average life), i.e. choosing a "Mean Return Period" $P_R$ of the earthquake using the following relation:

$$P_R = -Y / \ln(1 - P_{EY}) \qquad (1)$$

An example of the mean return period associated to each limit state for ordinary buildings ($Y$=50 years) is shown in Table 1 (ASCE 2006), although every country has its own values. In this way, acceptable structural performances are clearly defined and seem to be a rational way to approach the design (the lesser the earthquake intensity, the lesser the acceptable damage) (e.g. ASCE 41-13 Table C2-4). Nevertheless, there are some critical points of the standard procedure that should be highlighted. The shaking parameter (usually the spectral acceleration) related to the mean return period $P_R$ is calculated using the Probabilistic Seismic Hazard Analysis approach (PSHA) (Cornell 1968).

Table 1. Typical mean return period

| Limit state | Probability of exceedance | Mean return period |
|---|---|---|
| OL | 50%/50 years | 72 |
| IO | 20%/50 years | 225 |
| LS | 10%/50 years | 475 |
| CP | 2%/50 years | 2475 |

Despite being widely used, this method has been strongly criticised by seismologists (e.g. Stein et al. 2012, Nekrasova et al. 2015 ), statisticians (e.g. Freedman et al. 2003, Castanos et al. 2002) and professionals (e.g. Rugarli 2014). The main criticisms are due to:
- Poor physical assumptions (e.g. poissonian occurrence of earthquakes, point-source hypothesis, "Mean Return Period" $P_R$ as if earthquakes could be considered periodic).
- Poor mathematical assumptions (e.g. confusing the probability of exceedance - a dimensionless quantity - with the rate of exceedance - a frequency; the two quantities can be equalized only for large numbers, and strong earthquakes do not satisfy this stringent requirement).
- Lack of reliable data, above all when treating strong earthquakes.
- Lack of validation of the results.
- Unrealistic intensity when using a small probability.

However, the scientific community did not reach a commonly accepted opinion and several papers have been written to support PSHA against those physically well rooted criticisms (e.g. Hanks et al. 2012, Iervolino 2013), creating an endless tit for tat that we do not want to enter here (for recent reviews see Panza et al. 2014).

In this paper we focus on what engineers should use as seismic input. Starting from the analysis of Equation 1, we will show that the use of PSHA to design a structure at any performance level that need not to be exceeded ever is erroneous, even if PSHA were absolutely accurate.

With this result at hand, in sections 3 and 4 we propose an alternative procedure to determine the seismic input based on a deterministic approach and the definition of the "Maximum Deterministic Seismic Input" (MDSI).

## 2 CONSIDERATION ON THE CURRENT PRACTICE OF SEISMIC INPUT DEFINITION AND PBD

The most advanced international seismic codes define the seismic input to assess structural performances as a function of:
- The importance of the structures (risk category);
- The performance structural level that has to be reached.

The importance of a structure is related with the hypothetical consequences of its failure (usually in terms of expected human losses).

This leads the codes to increase the expected structural performance with increasing importance of the structure. This process is represented in Figure 1.

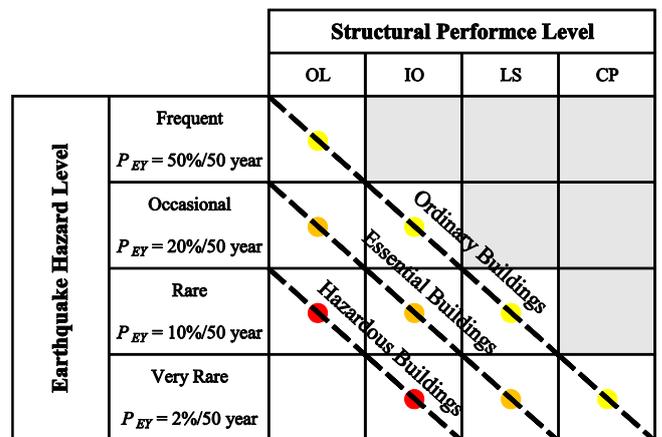

Figure 1. Standard Structural Performance Objectives

The change of the seismic input level with changing structural performances is something structural engineers should agree with. Indeed, we want better structural performances for earthquakes that occur frequently (i.e. for low intensities) and, on the other hand, we can accept high damages for a very rare earthquake. Nevertheless, the codes usually change the seismic input level not only as a function of the structural performance level that has to be reached, but also as a function of the importance of structures (risk category).

For example, in the ASCE 41-13 (ASCE 2013), ordinary structures should reach the basic safety objective. When designing new buildings this objective consists in achieving the Collapse Prevention level for an earthquake with $P_R$=2475 years (called BSE-1N) and the Life Safety level for an earthquake with $P_R$=475 years (called BSE-2N). When evaluating existing buildings, the Collapse Prevention level should be achieved for an earthquake with $P_R$=975 years (called BSE-1E) and the Life Safety level for an earthquake with $P_R$=225 years (called BSE-2E). Similarly, in the Italian Building Code (NTC08,

C.S.LL.PP. 2008) the building importance is taken into account by changing the reference average life of the structure, $Y$ in Equation 1. The higher the class of the structure, the higher its average life $Y$, which leads to an increase of the mean return period associated with the design seismic input. As a consequence, adopting the Italian Building Code, we should design a residential building that reaches the Collapse Prevention level for an earthquake with $P_R$=975 years (i.e $P_{EI}$=5% / $Y$=50 years) and, on the other way, an Essential Building (e.g. a school) should be designed to reach this level when subjected to an earthquake with $P_R$=1462 years (i.e. $P_{EI}$=5% / $Y$=75 years). These procedures should aim to optimize the economic effort to build structures in seismic areas. However, focusing on the Collapse Prevention level and assuming that the mean return period concept is reliable, this means that if an earthquake with $P_R$=1462 years happens, the residential building designed in accordance with the Italian Building Code would collapse. If an earthquake with $P_R$=2000 years happens, even a school would fall down. Is this reasonable? We believe not.

At a first glance these probabilities of occurrence could appear very low, but this objection is sitting on the erroneous and thus very misleading concept of mean return period. As it has been recently shown (Bizzarri et al., 2014) physical roots for $P_R$ are lacking and thus it represents a rather arbitrary choice and nothing more. Actually, events that have never happened before happen every day (Perrow 2011, Taleb 2007).

Analysing the earthquakes phenomena, the following considerations can be made:
– Every structure at a given site, regardless of its importance, experiences the same shaking when an earthquake of a given epicentral intensity/magnitude occurs;
– Nobody can tell with precision when an earthquake with a pre-fixed magnitude/intensity will happen;
– It is possible to estimate with some uncertainty the maximum magnitude/intensity that could affect a site.

These facts lead to state that choosing the seismic input as a function of the mean return period $P_R$ to optimize the cost might be acceptable, even if mathematically wrong, as long as the performance level we want to assess does not involve the failure of the structure (and $P_R$ could be varying from structure to structure). Indeed, PSHA gives structural engineers only a "vague idea" of how much an area is prone to earthquake phenomena, but nothing tells about the maximum earthquake that could happen and when it could happen.

When assessing the Collapse Prevention level, we are looking for a situation that could involve the loss of the structure. The failure of the structure means human losses and therefore, given the fact that an engineer cannot control the earthquakes phenomena, the least we can do is to use an "upper-bound" ground motion to design buildings at the CP level. As a rule, an "upper-bound" ground motion should be used to assess every structural performance which does not need to be exceeded ever (e.g. Immediate Occupancy level in an Hospital). To take into account the importance of the structure, engineers should increase or decrease the acceptable building performance while subjected to the same ground motions and do not reduce it, keeping in mind to assure at least the collapse prevention for the "upper-bound" ground motion. This is because engineers can govern the building performance through the designing procedure, but cannot predict future earthquakes.

Furthermore, this design procedure is supported by the evidence that the losses due to small and frequent earthquakes mainly involve non-structural element and content, while the losses due to an high intensity/magnitude earthquake are mainly in terms of structural damage and human lives. Usually, the overall cost of a building is subdivided in (Miranda et al. 2003):
– 8-18% of structural components
– 48-62% of non-structural components.
– 20-44% of contents.

Then optimizing costs using a probabilistic value of ground motions to assess the objective levels that involve non-structural and content damage may seem to be appropriate. When evaluating the collapse prevention level, the cost benefits due to a probabilistic decrease of ground motions involve a very small percentage of the overall cost in comparison with the cost of designing a structure for an "upper-bound" earthquake. Furthermore, collapse and human losses cannot be excluded. In other words, the game is not worth the candle.

3 SEIMIC HAZARD AND PBD, WHAT ENGINEERS SHOULD USE

Historically, a target probability of exceedance $P_{EY}$ of 10% in 50 years has been used worldwide as a reference to design ordinary buildings. These values of ground motion, as it could be expected given their probabilistic nature, have been being repeatedly exceeded (Kossobokov et al. 2012).

Moreover, the comparison between different probabilistic hazard maps reveals how the peak values (e.g. PGA with $P_{EY}$=10%/50 years) are not consistent from map to map, and large differences have been found (Nekrasova et al. 2015).

These observations and other engineering considerations (FEMA 2003) have led, in some countries, to a change of the value of $P_R$ from 10% to 2% in 50 years. Focusing on the considerations mentioned in section 1, to design or evaluate a building at the collapse prevention level, an input that cannot be exceeded should be used. If the probabilistic method were reliable, this "safety" level of ground motion should be calculated for a mean return period equal to the limit of Equation 1 as $P_{EY}$ approaches zero. However, evidence shows that a high increase of the mean return period $P_R$ often results in unreasonable high values of ground motion intensities, in particular in low-seismicity areas (Andrews et al. 2007, Klügel 2005, Bommer et al. 2004).

To avoid this limit, probabilistic maps have been complemented by the use of deterministic maps. For example the FEMA P-750 (FEMA 2009), as well as ASCE 7-10 (ASCE 2010), defines the "risk-targeted Maximum Considered Earthquake" ($MCE_R$) as the minimum between probabilistic and deterministic ground motions.

Deterministic Seismic Hazard Assessment (DSHA) is usually a scenario based approach where the hazard is chosen as the maximum ground motion of a set of individual earthquakes (magnitude and distance) that could happen at a site. The reason for using deterministic spectral accelerations as a cap for the probabilistic values, as written in FEMA P-750, is that "*deterministic ground motions provide a reasonable and practical upper-bound to design ground motions*". Anyway, remembering the considerations of section 2, when evaluating a structure at the Collapse limit prevention a probabilistic reduction of the seismic ground motion is not reliable, while the consequent cost reduction is a small percentage of the overall cost of the building. This should lead engineers to design at the collapse limit state using the aforementioned "upper-bound" ground motion. This means, as reported in FEMA P-750, using an input based on a Deterministic Seismic Hazard Assessment (DSHA). Hence, from now on let us define this level of ground motion as the "Maximum Deterministic Seismic Input" (MDSI).

### 3.1 Seismic input definition: a proposal for the MDSI

The use of deterministic methods is considered the most reliable way to calculate the Seismic Collapse Input. However, this statement is rather vague since several deterministic methods are available. Usually, standard DSHAs rely on the use of Ground Motion Prediction Equations (GMPE). These consist in empirical relations, and relative uncertainties, based on observations. Therefore, using standard DSHA approaches, the seismic input is defined as a fixed percentile (i.e. $84^{th}$ percentile) ground motion for a range of characteristic earthquakes. However GMPEs are affected by some severe limitations, namely:
- strong dependence on available data, which are usually limited;
- disruption of the tensor nature of earthquake phenomena;
- time history ground motions cannot be obtained (i.e. only spectral quantities can be handled);
- the effects due to the complexity of source rupture (i.e. directivity) can hardly be taken into account;
- local effects cannot be included in the analysis, since they are not persistent but earthquake source dependent (Molchan et al. 2011).

To overcome these drawbacks an upgraded deterministic approach has been developed, called Neo Deterministic Seismic Hazard Assessment (NDSHA) (Panza et al. 2001, Panza et al. 2012). NDSHA does not use GMPE. Instead, this is a scenario based procedure which supplies realistic time history ground motions calculated as the tensor product between the earthquake source tensor and the Greens function of the medium. It is based on seismic-wave propagation modelling starting from the knowledge of the seismic sources and the structural properties of the earth. NDSHA allows to take into account the complexity of the source process, as well as the site effects. Peak values of ground displacement, velocity and acceleration, as well as response spectra are defined by means of envelops of a large number of earthquakes that can occur at a given site. From an engineering point of view, seismograms provided by NDSHA (e.g. Figure 6) simulations also allow to run time history analysis using site specific inputs even where no registrations are available.

So far it has been applied in several countries at different levels of detail (e.g. Panza et al. 2012).

To define the MDSI, here we propose a standardization of the NDSHA procedure to fit the needs of engineers. The procedure has been applied to the Italian territory.

As a first step, a "Regional Scale Analysis" (RSA) is carried out using a large number of possible sources. It provides the "Maximum Deterministic Seismic Input at bedrock" ($MDSI_{BD}$), without considering the site effects (see section 3.1.1). The $MDSI_{BD}$ could be used, if corrected by means of the standard approximate soil coefficients (e.g. prospects 3.2 and 3.3 from EC8-1), for a standard design for ordinary buildings and a preliminary design for hazardous buildings. The RSA is then used as a reference to choose the most dangerous sources for the site of interest. With reference to these sources, a detailed "Site Specific Analysis" (SSA) which takes into account the local structural heterogeneities, is then carried out for each source-to-site path. The SSA allows to determine the "Maximum Deterministic Seismic Site Specific Input" ($MDSI_{SS}$). The $MDSI_{SS}$ (see section 3.1.2) is used to design the structures at the Collapse Prevention level or at any other level of performance that must not be exceeded.

*3.1.1 Regional Scale Analysis (RSA) and $MDSI_{BD}$*

At regional scale, synthetic seismograms are computed with the modal summation technique (Panza 1985) in far field conditions and with the matrix impedance method (Pavlov 2009) in near field conditions. As it is well known, to perform NDSHA the properties of the sources and structural models of the earth are needed. As a rule, NDSHA allows us to consider all the available information about the spatial distributions of the sources, their magnitudes and focal mechanisms, as well as about the properties of the inelastic media crossed by earthquake waves. For example, as far as the Italian territory is concerned, the potential sources are defined combining the parametric catalogue of Italian earthquakes CPTI04 (Gasperini et al 2004), the earthquakes catalogues for Slovenia and Croatia (Zivcic et al. 2000, Markusic et al. 2000), the ZS9 seismogenic zones (Meletti et al. 2004) and the associated focal mechanisms. In addition to such information, seismogenic nodes (i.e. zones prone to strong earthquakes) have been identified through a morphostructural analysis (Gorshkov et al. 2002, 2004, 2009). The nodes have been represented as circles of radius R=25 km within which earthquakes have magnitude $M_N \geq 6$ or $M_N \geq 6.5$. Possible epicentres over the territory are discretized into 0.2°x0.2° cells and the maximum recorded magnitude within them is used for computations. The spatial uncertainties and the source dimensions are taken into account applying a smoothing procedure (Panza et al. 2001) that takes roughly into account both location uncertainties and source dimensions. Once the smoothing procedure has been completed, only the sources falling into the seismogenic zones and into the seismogenic nodes are retained. The magnitude of each cell is then set equal to the maximum among: a) the magnitude $M_N$ of the seismogenic nodes, b) the magnitude resulting from the smoothing procedure, c) a minimum magnitude of 5. The resulting map of seismic sources for the Italian territory is shown in Figure 2 (Panza et al. 2012). On account of the fact that the root mean square (r.m.s.) of magnitude global determinations varies from about 0.2 to 0.3 (e.g. Bath 1973, Petersen at al. 2014), uppermost values corresponding to 2 r.m.s. can be easily obtained adding 0.5 magnitude units to the magnitude values reported in the catalogues. Thus, this approach envelops uncertainties rather than trying to quantify them probabilistically.

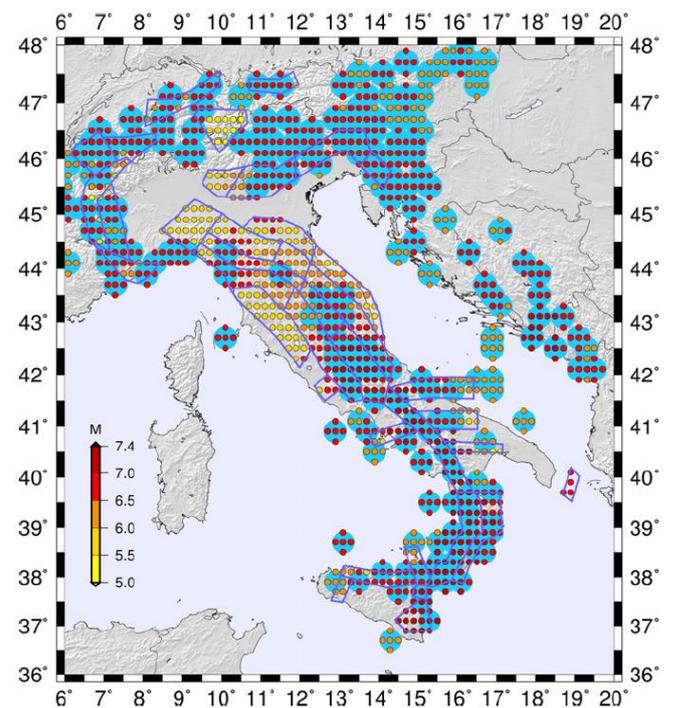

Figure 2. Map of the seismic sources for the Italian territory. The blue areas represent the seismogenic nodes (Panza et al. 2012)

These sources are modeled as size- and time-scaled point sources (STSPS). A double-couple that represents a focal mechanism consistent with the tectonic character of the seismogenic zone is placed at the centre of each cell. The depth is chosen as a function of the magnitude (10 km for $M \leq 7$, 15 km for $M > 7$). The moment-magnitude relation is chosen as that given by Kanamori

(Kanamori 1977) while the dimension of the sources is accounted for using the spectral laws proposed by Gusev (Gusev 2011, personal communication).

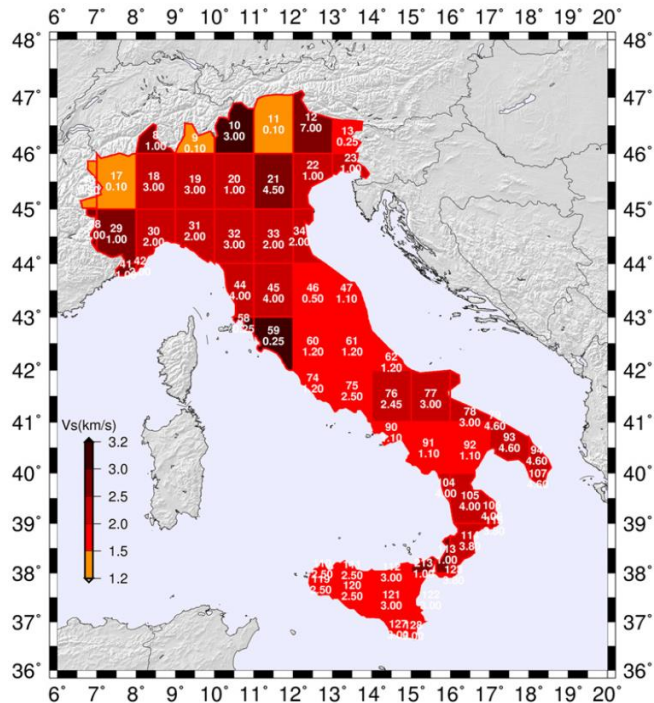

Figure 3. Cellular structural model (Panza et al. 2012); the S-wave velocity (color) and the thickness in kilometers (bottom number) of the topmost layer of each structure are represented

The physical properties of the source-site paths are defined using a set of cellular structures (Figure 3) obtained through an optimized nonlinear inversion of surface wave dispersion curves (Brandmayr et al. 2010).

To account statistically for the variability of the ground motion at a site due to the rupture process and the consequent directivity effects, 100 different models of rupture are generated, for each source-to-site path, using the PULSYN algorithm (Gusev 2011). Synthetic seismograms are then computed at each node of a grid of 0.2°x0.2° over the Italian territory at a frequency up to 10 Hz.

In particular, to define the $MDSI_{BD}$, for each source within 150 km from the site of interest, the response spectra corresponding to the median, $85^{th}$ percentile and $95^{th}$ percentile are computed (no normal or other questionable fixed distributions are considered, i.e. 100 different response spectra are derived from synthetic seismograms for each source, then for each period the values which have been exceeded in the simulations 50 times, 15 times and 5 times are recorded). An example of this procedure, for one source, is shown in Figure 4.

Once these spectra have been computed for all the sources, the envelope of the median spectra is calculated and the associated $85^{th}$ and $95^{th}$ percentile are reported.

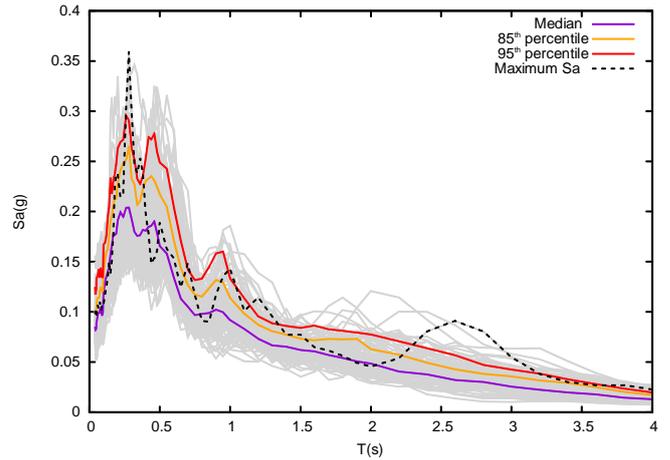

Figure 4. Variability of the response spectra at a site as obtained considering 100 different models of rupture

The sources that contribute to the definition of the envelope of the median spectra at the bedrock (Figure 8), could be used as scenarios to run a Site Specific Analysis (SSA). The last step is to represent the median, $85^{th}$ and $95^{th}$ response spectra using a shape in compliance with the Italian code (NTC08), function of the parameters $T_c^*$, $F_0$ and $a_g$. This is achieved by minimizing the difference between the real shape and the standard shape (Figure 9).

The Maximum Seismic Input at the bedrock ($MDSI_{BD}$) is then set equal to the $95^{th}$ percentile. This choice might look wrong, given the fact that using the $95^{th}$ percentile means that 5 computed signals out of 100 have an higher spectral acceleration. However it can be explained by Figure 4, where the $95^{th}$ percentile for one source and response spectra with the maximum pseudo acceleration are highlighted. As it can be seen, the signal with the highest spectral accelerations has a peak value at about 0.35s, but for the other periods the spectral accelerations are less than those of the $95^{th}$ percentile response spectra. This happens for all the response spectra that give the maximum spectral acceleration at each period. Hence, the use of the envelope of the maximum spectral accelerations at each period as input for a response spectrum analysis would lead to oversizing the structure to be designed.

In fact, the $95^{th}$ percentile response spectra computed as proposed represents an envelope of all the simulations, and it takes into account the highest values at each period without considering the effects of the isolated peaks.

### 3.1.2 Site Specific Analysis (SSA) and $MDSI_{(SS)}$

The $MDSI_{BD}$ obtained by the RSA proposed in section 3.1.1, as the name implies, is valid since the site of interest is placed on a bedrock soil.

This condition is quite rare and usually the ground motion at a site is strongly dependent on the interaction between source radiation and lateral heterogeneities, whether topographical or due to the presence of soft-sedimentary soil.

As a rule, local "amplifications" are evaluated in a simplified manner by modifying the shape of the response spectra at the bedrock using different coefficients. These coefficients are function both of the surface layer and of the topographical condition. A more detailed computation of the local amplifications might be carried out using the ratio between the horizontal and the vertical response spectra (H/V ratio) (Nakamura 1989). This widely used factor is obtained from seismic noise, assuming that the vertical ground motion is not affected by the superficial layer. Anyway, this method has been demonstrated to be unable to give correct amplifications, as well (Panza et al. 2012). In fact, the vertical component of motion can be amplified by local conditions, too. In particular, amplifications of both vertical and horizontal components of motion are strongly dependent not only from the soil and topography characteristics, but also from the incidence angles of the radiated wavefield.

To overcome these limits, a method based on computer simulations that exploit the knowledge about the source process, the path source-to-site and the site conditions has been developed. This hybrid method combines the modal summation and the finite-difference technique (Faeh et al. 1994). The wave-field generated by the modal summation technique is introduced in the mesh that defines the local heterogeneous area and it is propagated according to the finite-differences scheme shown in Figure 5 (Panza et al. 2012).

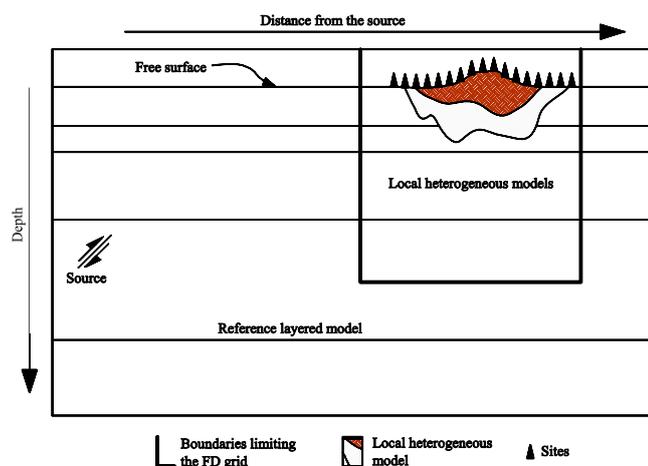

Figure 5. Schematic diagram of the hybrid method (Panza et al. 2001)

To define the Maximum Deterministic Site Specific Input ($MDSI_{SS}$) the same procedure of section 3.1.1 can be adopted, adding the local heterogeneous model. To reduce the time costs, the procedure is applied only for the sources (for the path source to site) that give the largest bedrock hazard, that is for the scenarios that contribute to develop the envelope of the response spectra (see Figure 8). In other words, the Site Specific Analysis (SSA) is a RSA carried out only for the most hazardous sources for the site of interest but taking into account the local conditions. Accordingly with the results of section 3.1.1, the $MDSI_{SS}$, as for a RSA, is set equal to the 95$^{th}$ percentile response spectra.

From an engineering point of view, in addition to accounting for realistic site amplifications, a SSA provides realistic and site specific synthetic seismograms (Figure 6). This feature is truly important given the fact that the number of recorded ground motion is very low, particularly for large earthquakes. A preliminary Site Specific Analysis is then essential to run time history structural analysis using seismograms representative of the dynamic characteristics of the site of interest.

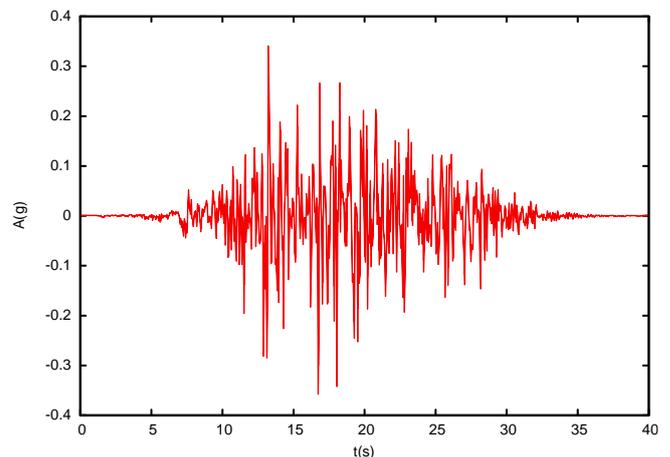

Figure 6. Example of synthetic seismogram generated by means of NDSHA method

### 3.2 Application to Performance Based Design

The most advanced design patterns consider the possibility to assess different levels of performance, namely the usual Operational level OL, Immediate Occupancy IO, Life Safety LS and Collapse Prevention CP. These levels are by definition function of the damage that is accepted to occur in the structural and non-structural element of a building when subjected to a certain level of ground motion. The level of damage is a function of the structure and it is usually defined in terms of acceptable storey drift or acceptable rotation in the plastic hinges. Instead, the level of ground motion associated to them (the ground motion used to check if the level has been reached), is usually taken into account using a probabilistic approach. In the light of the considerations made so far, this process needs a

review. In particular, the Performance Based Design approach should follow these steps:
1. Choose a risk category for the building (namely its importance, e.g. Ordinary Building, Essential Building or Hazardous Building);
2. Choose the *target performance level* (TPL) that is needed not to be exceed ever (the worst acceptable performance the structure should exhibit during its life), that is the structural performance level associated with the $MDSI_{SS}$;
3. Choose, as a consequence of the TPL chosen in step 2, the *lower performance levels* (LPL) and the associated ground motions (see Figure 7).

The difference between this approach and what is usually done by the codes (e.g. Italian Building Code) is that we take into account the importance of the structure (its risk category) by changing only the structural performance level to be checked, and not changing the seismic input. Furthermore, the worst performance the structure is allowed to reach is always associated with the $MDSI_{SS}$ level of hazard.

Regarding step number three, the definition of the ground motion is necessary. LPLs are, by definition with respect to the TPL, levels of performance that could be exceeded during the life of the structure and usually less important than TPL. This means that the spectral level associated with LPL has to be less than $MDSI_{SS}$ and, given their conventional nature, several ways could be followed. In first approximation the use of probabilistic values, rather than using a "rule of thumb", may be acceptable even though it is based on the non-physically routed concept of return period. Alternatively and equally arbitrarily, such levels could be defined as a fraction of $MDSI_{SS}$ response spectra (for example 2/3 of $MDSI_{SS}$ for medium seismic input level and 2/5 of $MDSI_{SS}$ for low seismic input level). This procedure is summed up in Figure 7, whereby two probabilistic values are suggested as examples (once $MDSI_{SS}$ have been defined, the assumption of the seismic input associated with LPLs are more an economical choice rather than an engineering one).

With regards to the Structural performance level, it should be noted that the Life Safety level should be intended as preserving the life of the structure (reparability), rather than preserving human safety (usually assured by checking the CP). In other words, it should be assessed as a Still Reparable (SR) structural level.

As an example of this procedure, a residential building should be designed at the CP level for the $MDSI_{SS}$ seismic input, while at the SR and IO levels a reduced seismic input could be used.

Figure 7. Proposed PBD procedure considering the MDSI

An essential building should be designed for the $MDSI_{SS}$ at the SR, while the IO and OL levels should be assessed with a lower value of the seismic input. Of course, by using a linear analysis, there is the necessity to define a different behaviour factor for the SR and CP levels, given the fact that these performance levels involve non-linear behaviour and different percentage of acceptable damage. It is important to highlight that the uncertainties, both structural and related to the seismic input, are such that it is impossible to predict exactly the seismic behaviour of a structure and this is the main aspect that should lead engineers to design by means of envelops instead of using useless complicated probabilistic calculations.

Clearly, the procedure proposed is conventional and should be used as a minimum requested performance to assess the building during its design stage and to assure a reasonable level of safety.

4 $MDSI_{BD}$ EXAMPLES

In this section, examples of the definition of the $MDSI_{BD}$ acceleration response spectra are provided and compared with the acceleration response spectra given by the Italian code. In particular a RSA has been carried out, as explained in section 3.1.1, for three different cities: L'Aquila, Trieste and Gorizia. Figure 8 shows the envelope of the median spectrum, and the associated $85^{th}$ and $95^{th}$ percentile, of all the sources within a distance of 150 km from the site of L'Aquila. As it can be seen the most dangerous sources are number 15, 16 and 21, which are used to run a SSA and find the $MDSI_{SS}$.

Figure 9 shows, again for the site of L'Aquila, the fit between the real spectrum shape and the shape given by the Italian Building Code (NTC08). This step is carried out mainly to reduce the parameters needed to define the spectrum and to allow for a comparison with the probabilistic values at bedrock given by the code.

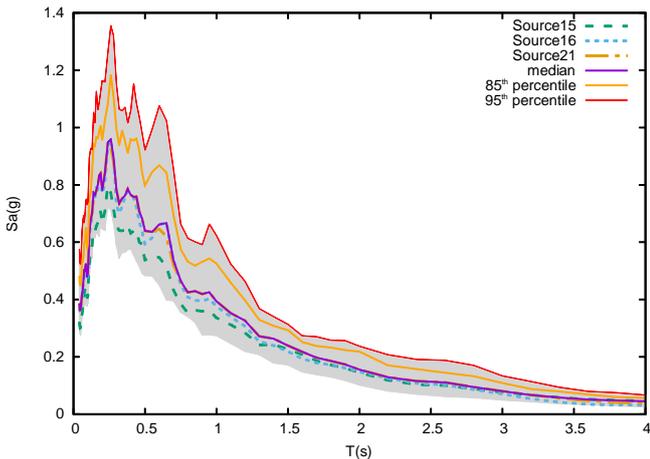

Figure 8. Envelope of the median response spectra, and the associated 84[th] and 95[th] percentile, of the sources within a distance of 150 km from the site of L'Aquila.

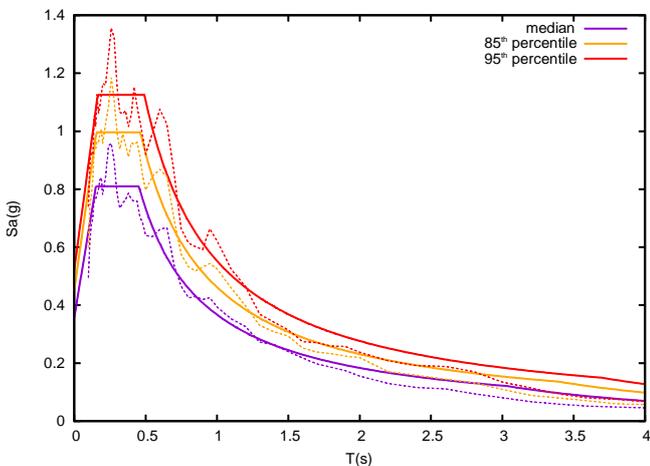

Figure 9. Site of L'Aquila: Fitting between the real spectrum shape and NTC08 shape (C.S.LL.PP. 2008)

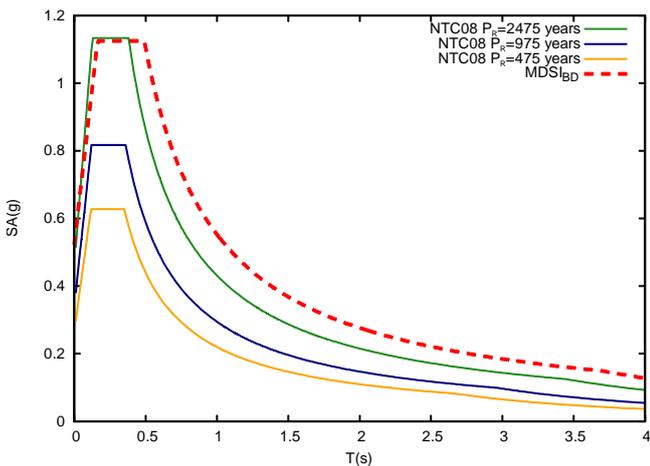

Figure 10. Site of L'Aquila: Comparison between the $MDSI_{BD}$ and the NTC08 probabilistic response spectra

Figures 10, 11 and 12 show, for the three sites considered, a comparison between the $MDSI_{BD}$, the maximum seismic input provided by the code ($P_R$=2475 years), the input actually associated by the code to the collapse prevention structural level for a standard residential building ($P_R$=975 years), and the response spectra usually adopted to design standard building at life safety level ($P_R$=475 years). As it can be seen, the $MDSI_{BD}$ response spectra, which should be a cap for the spectral accelerations, are very close to the response spectra with $P_R$=2475 years.

Looking at Figures 10, 11 and 12, the spectral accelerations for the $MDSI_{BD}$ spectra are directly comparable with that given by the $P_R$=975 years response spectra, given the fact that they are related to the same structural performance level. Instead, to compare the $MDSI_{BD}$ response spectra with the $P_R$=475 years spectrum, a reduction factor should account for the different level of acceptable damage associated with them. This reduction factor could be considered equal to the ratio between the behaviour factor (response coefficient) $q_{CP}$ at the CP level and the behaviour factor $q_{LS}$ at the LS level, which is usually the one provided by the codes. It varies from structure to structure, but on average it could be set equal to 1.3÷1.5.

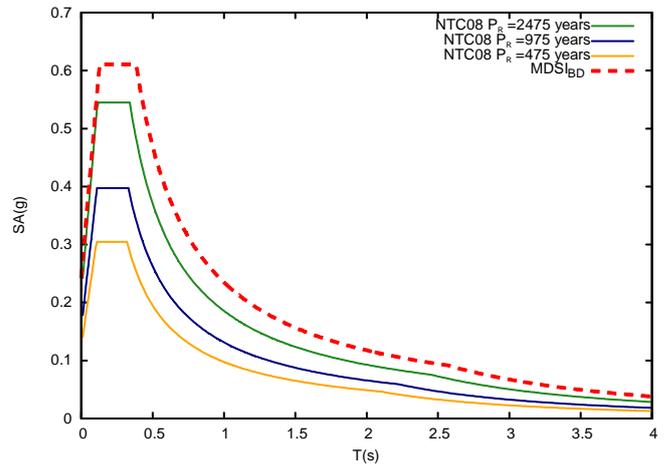

Figure 11. Site of Trieste: Comparison between the $MDSI_{BD}$ and the NTC08 probabilistic response spectra

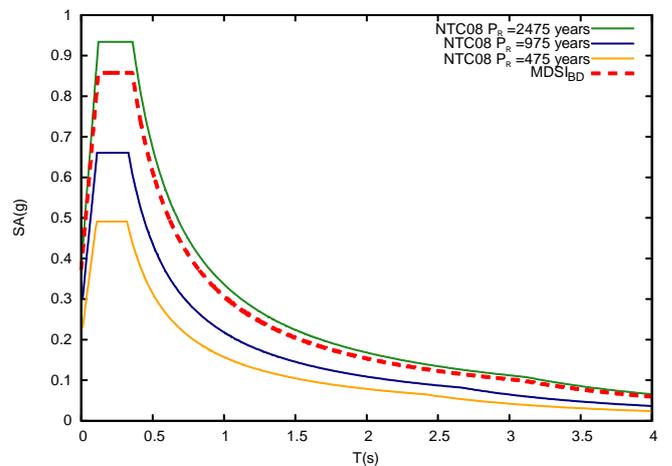

Figure 12. Site of Gorizia: Comparison between the $MDSI_{BD}$ and the NTC08 probabilistic response spectra

Considering this reduction factor, the $MDSI_{BD}$ are in all the cases considered higher than the $P_R$=475 years response spectra used in the design of residential buildings, but not so much as to suggest a significant increase in costs.

The characteristic spectral parameters necessary to define the $MDSI_{DB}$ response spectra in compliance with the shape define by the Italian Bulging Code are shown for the cities of L'Aquila, Trieste and Gorizia in Table 2.

Table 2. Characteristic spectral parameter for the considered sites

| Site | Response Spectra | $a_g$ [g] | $F_0$ | $T_c^*$ [s] |
|---|---|---|---|---|
| L'Aquila | 50th percentile | 0.36 | 2.26 | 0.45 |
| L'Aquila | 85th percentile | 0.45 | 2.22 | 0.46 |
| L'Aquila | $MDSI_{BD}$ (95th percentile) | 0.52 | 2.15 | 0.49 |
| Trieste | 50th percentile | 0.17 | 2.47 | 0.36 |
| Trieste | 85th percentile | 0.21 | 2.44 | 0.39 |
| Trieste | $MDSI_{BD}$ (95th percentile) | 0.24 | 2.51 | 0.38 |
| Gorizia | 50th percentile | 0.26 | 2.28 | 0.33 |
| Gorizia | 85th percentile | 0.32 | 2.30 | 0.35 |
| Gorizia | $MDSI_{BD}$ (95th percentile) | 0.36 | 2.30 | 0.36 |

## 5  CONCLUSIONS

In this paper, starting from a review of the standard Performance Based Design procedure, the limits of the probabilistic seismic input definition to asses any structural performance that should not be exceeded in a building have been shown. To overcome these limits, a new performance based design strategy has been proposed, whereby the seismic input is calculated using the Neo Deterministic Seismic Hazard Assessment (NDSHA). By means of envelopes on a wide range of NDSHA simulations which take into account for the seismic history, the seismogenic zones and the seismogenic nodes (zone prone to strong earthquakes), the "Maximum Deterministic Seismic Input" (MDSI) has been defined. MDSI can be defined at bedrock ($MDSI_{BD}$) using a Regional Scale Analysis (RSA) and at the free surface ($MDSI_{SS}$) using a Site Specific Analysis (SSA) or, for a preliminary design, using the standard approximate soil coefficients (e.g. prospects 3.2 and 3.3 from EC8-1). A SSA provides realistic site specific seismograms that are useful to run time history analysis even where no registrations are available.

$MDSI_{SS}$ is always associated with the worst structural performance acceptable for a building, called *Target Performance Level* (TPL). In this way, the importance of the structure (risk category) is taken into account by changing the structural performance level to be checked, and not the seismic input. The performance levels that involve less percentage of damage with respect to the TPL are called *Lower Performance Levels* (LPL). Given its conventional nature, the seismic input level associated to the LPLs can be found either using probabilistic values or reducing the $MDSI_{SS}$ spectral accelerations.

As regards to ordinary buildings, the increase of demand due to the use of the MDSI response spectra may be reduced increasing the behaviour factor, since it is linked to a different limit state. Comparisons between the proposed procedure and the response spectra with $P_R$=475 years provided by the Italian Code have shown that the use of the MDSI should not lead to a significant increase in costs. This procedure should be used as a minimum requested performance. Moreover, being the MDSI independent from the importance of the structure, by adopting this approach it is possible to identify the margin of safety with respect to a TPL, which can be used to classify the seismic vulnerability of structures in a comparable way from building to building.

ACKNOWLEDGEMENT

We would like to thank Paolo Rugarli for his helpful comments. This paper has been supported by the "*Fondazione Cassa di Risparmio di Gorizia*" and FRA 2014 - University of Trieste.